\documentclass[article,11pt]{memoir}
\pdfoutput=1
\usepackage[utf8]{inputenc}
\usepackage[english]{babel}
\usepackage[scaled]{beramono}
\usepackage[T1]{fontenc}
\usepackage[left=3.3cm, right=3.3cm, top=3cm]{geometry}
\usepackage{changepage}
\usepackage{amsmath,amssymb,bm,mathtools,dsfont,upgreek,stmaryrd}
\usepackage{kpfonts}
\usepackage{shadethm}
\usepackage{pdfpages}
\usepackage{enumitem}
\usepackage{array,booktabs}
\usepackage{graphicx}
\usepackage{float}
\usepackage{hyperref}
\usepackage{tikz}
\counterwithout{section}{chapter}
\author{August Bjerg}

\newcommand\SetSymbol[1][]{\mathrel{#1\vert}}
\DeclarePairedDelimiterX\Set[1]\{\}{%
	\def\given{\SetSymbol[\delimsize]}%
	#1}
\newcommand{\norm}[1]{\ensuremath{\lVert #1 \rVert}}
\newcommand{\abs}[1]{\ensuremath{\lvert #1 \rvert}}
\newcommand{\Indre}[2]{\ensuremath{\langle #1,#2 \rangle}}

\usepackage{xcolor}
\usepackage[ntheorem]{empheq}
\usepackage[ntheorem,framemethod=tikz]{mdframed}
\usepackage[amsmath,  thmmarks]{ntheorem}
\definecolor{shadecolor}{gray}{1}
\definecolor{rulecolor}{gray}{0.7}
\mdfdefinestyle{thm}{ 
	usetwoside=false,
	skipabove=1em plus 0.4em minus 0.6em,
	skipbelow=0.5em plus 0.4em minus 0.2em,
	innerleftmargin=.5em,
	innerrightmargin=.5em,
	innertopmargin=0.5em,
	innerbottommargin=0.5em,
	leftmargin=\dimexpr-.5em-1pt\relax,
	rightmargin=\dimexpr-.5em-1pt\relax,
	linewidth=1pt, 
	linecolor=rulecolor, 
	backgroundcolor=shadecolor, 
	splittopskip=1.2em minus 0.2em,
	splitbottomskip=0.5em plus 0.2em minus 0.1em,
}

\theoremseparator{.}

\newmdtheoremenv[style=thm,nobreak=true]{thm}{Theorem}
\newmdtheoremenv[style=thm]{thmB}[thm]{Theorem}

\newmdtheoremenv[style=thm,nobreak=true]{lemma}[thm]{Lemma}
\newmdtheoremenv[style=thm]{lemmaB}[thm]{Lemma}

\newmdtheoremenv[style=thm,nobreak=true]{pro}[thm]{Proposition}
\newmdtheoremenv[style=thm]{proB}[thm]{Proposition}

\newmdtheoremenv[style=thm,nobreak=true]{corollary}[thm]{Corollary}
\newmdtheoremenv[style=thm]{corollaryB}[thm]{Corollary}

\newmdtheoremenv[style=thm,nobreak=true]{definition}[thm]{Definition}
\newmdtheoremenv[style=thm]{definitionB}[thm]{Definition}

\theorembodyfont{\normalfont}

\theorembodyfont{\normalfont}

\theorembodyfont{\normalfont}
\theoremsymbol{\rule{7pt}{7pt}}
\newtheorem{remark}[thm]{Remark}

\theorembodyfont{\normalfont}
\theoremsymbol{\rule{4pt}{4pt}}

\theorembodyfont{\normalfont}
\theoremsymbol{ \ensuremath{\scriptstyle\blacktriangle}}
\newtheorem{example}[thm]{Example}

\theoremstyle{nonumberplain}
\theoremheaderfont{\itshape} 
\theorembodyfont{\normalfont} 
\theoremseparator{.} 
\theoremsymbol{\enspace\ensuremath\square} 
\newtheorem{proof}{Proof}

\usepackage[square,sort,comma,numbers]{natbib}
\newcommand*{\doi}[1]{DOI: \href{http://dx.doi.org/#1}{\texttt{#1}}}

\begin{document}
\begin{center}
\textbf{\LARGE{Convergence of operators with deficiency indices $(k,k)$ and of their self-adjoint extensions}}
\vspace{0.2cm}
\begin{center}
August Bjerg\let\thefootnote\textdagger \footnote{\href{mailto:aabjerg@math.ku.dk}{\texttt{aabjerg@math.ku.dk}}}
\\ \vspace*{.2cm}
Department of Mathematical Sciences, University of Copenhagen,
\\
Universitetsparken 5, DK-2100 Copenhagen Ø, Denmark
\end{center}
\renewcommand*\abstractname{Abstract:}
\abstractrunin
\vspace{0.5cm}
\begin{abstract}
We consider an abstract sequence $\{A_n\}_{n=1}^\infty$ of closed symmetric operators on a separable Hilbert space $\mathcal{H}$. It is assumed that all $A_n$'s have equal deficiency indices $(k,k)$ and thus self-adjoint extensions $\{B_n\}_{n=1}^\infty$ exist and are parametrized by partial isometries $\{U_n\}_{n=1}^\infty$ on $\mathcal{H}$ according to von Neumann's extension theory. Under two different convergence assumptions on the $A_n$'s we give the precise connection between strong resolvent convergence of the $B_n$'s and strong convergence of the $U_n$'s.
\end{abstract}
\end{center}
\section{Introduction}\label{sec1}
We investigate in the following the notion of strong resolvent convergence of sequences of self-adjoint extensions of already specified (unbounded) closed symmetric operators on a Hilbert space. For the general theory on these topics we refer to \cite{RS1} VIII and \cite{RS2} X and introduce now the framework in which we will be working for the present section as well as for Section \ref{sec3} where our main results are found. In Section \ref{sec2} we treat also more general operators than considered here.

Consider a symmetric and closed operator $A$ on an infinite dimensional separable Hilbert space $\mathcal{H}$\footnote[1]{We adopt the convention that the inner product on $\mathcal{H}$ is linear in the second entry} defined on a dense subspace $D(A)$. The kernels $\mathcal{H}_\mp:= Z(A^*\pm i)$ are the \emph{deficiency subspaces} and the pair $(\dim\mathcal{H}_+,\dim\mathcal{H}_-)$ is the \emph{deficiency indices}. We assume that the latter are equal and finite, i.e. $(\dim\mathcal{H}_+,\dim\mathcal{H}_-)=(k,k)$ for some $k=1,2,\dots$ (however, see Remark \ref{Infinity}). This implies, cf. \cite{RS2} Theorem X.2, that $A$ has self-adjoint extensions, and moreover any self-adjoint extension $B$ of $A$ is given by the rule
\[
D(B)=\Set{\phi_0+\phi_++U\phi_+\given\phi_0\in D(A),\; \phi_+\in\mathcal{H}_+},
\]
\[
B(\phi_0+\phi_++U\phi_+)=A\phi_0+i\phi_+-iU\phi_+
\]
where $U\colon\mathcal{H}_+\to\mathcal{H}_-$ is a unitary map which can be extended to a partial isometry on all of $\mathcal{H}$ by letting $U\phi=0$ for $\phi\in[\mathcal{H}_+]^\perp$. Conversely, all extensions of $A$ of this form are self-adjoint.

We introduce now sequences $\{A_n\}_{n=1}^\infty$ and $\{B_n\}_{n=1}^\infty$ of such operators. That is, the $A_n$'s are densely defined, symmetric and closed operators on $\mathcal{H}$ with deficiency subspaces $\mathcal{H}_\pm^n$ and deficiency indices $(k,k)$ independent of $n$, and $B_n$ is a self-adjoint extension of $A_n$ defined by a unitary map $U_n\colon\mathcal{H}_+^n\to\mathcal{H}_-^n$ (which can all, once again, be considered as partial isometries on $\mathcal{H}$) as described above for each $n$. In this set-up we think of $A$, $B$ and $U$ as limiting operators of the sequences of $A_n$'s, $B_n$'s and $U_n$'s respectively, and our main goal will be to examine the interplay between the convergence of these sequences. A very natural question is for example whether we can obtain results along the lines of
\begin{equation}\label{result}
\text{\emph{"Suppose $A_n\to A$. Then $B_n\to B$ if and only if $U_n\to U$."}}
\end{equation}
Of course one needs here to specify which notions of convergences we involve in this statement for it to be mathematically interesting. For the purposes of this note we focus on strong convergence of operators on Hilbert spaces. Hence, $U_n\to U$ should be understood as usual strong convergence of bounded operators and $B_n\to B$ as strong resolvent convergence of self-adjoint unbounded operators, i.e. as strong convergence of $(B_n+i)^{-1}$ towards $(B+i)^{-1}$ -- for an introduction to the topic and an explanation why this is in some sense the only "right" way of extending the concept of strong convergence to self-adjoint unbounded operators, see \cite{RS1} VIII.7. For the $A_n$'s, however, we cannot use this generalized version of strong convergence since these are not self-adjoint.

This issue will be addressed in Section \ref{sec2}. Once this theoretical framework is in place, we will gradually progress towards presenting statements of the form (\ref{result}) in Corollaries \ref{Cor1} and \ref{Cor2}. Finally, an exposition on the optimality of these results -- in particular of the latter -- is included for completeness.
\begin{example}\label{Motivation}
As a final note before diving into technical details we mention the structure of a class of motivational examples that illuminates why we even care to search for results like (\ref{result}).

Consider a sequence $\{\widetilde{A}_n\}_{n=1}^\infty$ of explicitly given symmetric differential operators on an open subset $\Omega$ of $\mathbb{R}^d$ defined on $D(\widetilde{A}_n)=C_c^\infty(\Omega)$. Now the usual way to realize $\widetilde{A}_n$ as a self-adjoint operator on $L^2(\Omega)=\mathcal{H}$ is the following: Let $A_n$ be the closure of $\widetilde{A}_n$ for each $n$ and if this is not already self-adjoint extend it by the above procedure to some self-adjoint operator $B_n$. Here we have an example where the sequence $\{A_n\}_{n=1}^\infty$ is concretely described and not often subject to change. It describes not only how the $A_n$'s but also (through the $A_n^*$'s) how the $B_n$'s act on their domain, and often it will not be to difficult to prove that $A_n\to A$ for some $A$ in an appropriate sense. We suppose that this convergence has been established. Moreover, natural examples of sequences of this form will in most cases satisfy the crucial property that all the operators have the same deficiency indices. The deficiency subspaces will be parts of solutions spaces of differential equations and usually the $U_n$'s will be simple maps between such spaces. Hence, in this case, strong convergence of the $U_n$'s is a property which is a lot easier to handle than the full strong resolvent convergence of the $B_n$'s.

Now one can envision a couple of situations: If a sequence of $B_n$'s is known, (\ref{result}) could help us determine a self-adjoint extension $B$ of $A$ so that $B_n\to B$ in the strong resolvent sense. One needs only to find the strong limit of the $U_n$'s (if this exists) and use this to extend $A$. If the strong limit of the $U_n$'s does not exists then the result will conversely tell us that the $B_n$'s do not converge towards any self-adjoint extension of $A$. On the other hand it could be that $B$ was a fixed self-adjoint extension of $A$ and the result could in the same manner be used to find a sequence of $B_n$'s which extends the $A_n$'s  and converge towards $B$ in the strong resolvent sense -- or whether such sequence exists at all.
\end{example}
\section{Strong graph convergence and convergence of graph projections}\label{sec2}
Now some candidates for types of convergences for the $A_n$'s in (\ref{result}) are treated. Along the way we introduce the machinery needed for both formulating and proving our main results. Firstly we need to introduce a particular notion of convergence of subspaces of a Hilbert space.
\begin{definition}\label{StrLimDef}
Let $\{V_n\}_{n=1}^\infty$ be a sequence of subspaces of a Hilbert space $\mathcal{H}$. The subspace
\[
V_\infty:=\Set[\Big]{x\in \mathcal{H}\given\substack{\text{There exists a sequence $\{x_n\}_{n=1}^\infty\subseteq\mathcal{H}$ with}\\ \text{$x_n\in V_n$ for each $n$ so that $x_n\to x$ as $n\to\infty$}}}
\]
is called the \emph{strong limit} of $\{V_n\}_{n=1}^\infty$ and we write $V_n\to V_\infty$ strongly.
\end{definition}
One should not be misled by the fact that we call this type of convergence "strong". We note that \emph{any} sequence of subspaces has a limit in the above sense (although it might be the trivial $0$-subspace), and hence this way of converging cannot be a particularly strong one. The word "strong" merely refers to the fact that $\{x_n\}_{n=1}^\infty$ should converge towards $x$ strongly, i.e. with respect to the Hilbert space norm.

Another notion of convergence of sequences of closed subspaces of a Hilbert space is that of the orthogonal projections onto these converging strongly towards the orthogonal projection onto a limiting subspace. In fact, this is generally a stronger notion of convergence of subspaces than the above "strong" convergence.
\begin{lemma}\label{SubSpaceLemma}
Let $\{V_n\}_{n=1}^\infty$ be a sequence of closed subspaces of a Hilbert space $\mathcal{H}$ and denote the orthogonal projections onto these by $\{P_n\}_{n=1}^\infty$. Denote similarly by $P$ the orthogonal projection onto another subspace $V\subseteq\mathcal{H}$.
\begin{itemize}
\item[(a)] $V$ is contained in the strong limit of $\{V_n\}_{n=1}^\infty$ if and only if $P_nx\to x=Px$ for all $x\in V$.
\item[(b)] If $P_n\to P$ strongly then $V$ is the strong limit of $\{V_n\}_{n=1}^\infty$.
\end{itemize}
\end{lemma}
\begin{proof}
(a): Assume on the one hand that $V$ is contained in the strong limit of $\{V_n\}_{n=1}^\infty$. Then, for any $x\in V$, there exists a sequence $\{x_n\}_{n=1}^\infty\subseteq\mathcal{H}$ with $x_n\in V_n$ for all $n$ so that $x_n\to x$. Hence, $\norm{P_nx-x}\leq\norm{x_n-x}\longrightarrow0$ as needed. The other implication is clear if one considers the sequence $\{P_nx\}_{n=1}^\infty$ for each $x\in V$.

(b): Assume $P_n\to P$ strongly and denote by $V_\infty$ the strong limit of $\{V_n\}_{n=1}^\infty$. By (a) we need only to argue that $V_\infty\subseteq V$ or equivalently $V^\perp\subseteq V_\infty^\perp$. However, if $y\in V^\perp$ then for any $x\in V_\infty$ we can choose a sequence $\{x_n\}_{n=1}^\infty\subseteq\mathcal{H}$ as for the $x$ in (a) and obtain
\[
\Indre{y}{x}=\lim_{n\to\infty}\Indre{(1-P_n)y}{x_n}=0
\]
proving $y\in V_\infty^\perp$ as needed.
\end{proof}
\begin{remark}\label{StrictlyStronger}
While Lemma \ref{SubSpaceLemma}(b) shows that convergence of projections is a stronger type of convergence than "strong" convergence in the sense of Definition \ref{StrLimDef}, the following example shows that it is actually \emph{strictly} stronger -- a fact which will be important later on.

Consider a sequence $\{V_n\}_{n=1}^\infty$ of subspaces of a Hilbert space $\mathcal{H}$ of the form $V_n=[\mathbb{C}x_n]^\perp$ where $x_n\in\mathcal{H}$ is of unit length and denote by $V_\infty$ the strong limit of this sequence. Suppose that $x_n=x_0$ is fixed for $n$ odd and $x_n=y_n$ for $n$ even where $\{y_n\}_{n=1}^{\infty}$ is a sequence which is weekly convergent towards $0$. Now $x_0\notin V_\infty$ since for $n$ odd we have $\operatorname{dist}(V_n,x_0)=1$. If, however, $x\in[\mathbb{C}x_0]^\perp$ then we can consider the sequence $z_n$ which is $x\in V_n$ for $n$ odd and $x-\Indre{y_n}{x}y_n\in V_n$ for $n$ even. As $\Indre{y_n}{x}\to0$ we see that $z_n\to x$ proving $x\in V_\infty$. We conclude that $V_\infty=[\mathbb{C}x_0]^\perp$.

On the the other hand the orthogonal projections $P_n$ onto the $V_n$'s do not converge strongly at all. In particular $P_nx_0$ is $0$ for $n$ odd and $x_0-\Indre{y_n}{x_0}y_n\to x_0$ for $n$ even.
\end{remark}
Letting operators once again enter the picture we can now easily define a notion of convergence of any sequence of operators on a Hilbert space: The strong graph convergence which is also treated in \cite{RS1} VIII.7.
\begin{definition}
Let $\{A_n\}_{n=1}^\infty$ be any sequence of operators on a fixed Hilbert space $\mathcal{H}$. If the graphs $\operatorname{Gr}(A_n)$ converge strongly towards the graph $\operatorname{Gr}(A)$ of some operator $A$ on $\mathcal{H}$ as subspaces of $\mathcal{H}\oplus\mathcal{H}$ then we say that $A$ is the \emph{strong graph limit} of the $A_n$'s and write $A=\operatorname{str.gr.lim }A_n$.
\end{definition}
Let us return to the case of a sequence of densely defined and closed operators $\{A_n\}_{n=1}^\infty$ for the remaining part of the section and fix once and for all the following convenient notation: By $\Gamma_\infty$ we mean the strong limit of $\{\operatorname{Gr}(A_n)\}_{n=1}^\infty$ and by $\Gamma_\infty^*$ the strong limit of $\{\operatorname{Gr}(A_n^*)\}_{n=1}^\infty$. Note that $(\phi,\psi)\in\Gamma_\infty$ if and only if there exists a sequence $\{\phi_n\}_{n=1}^\infty\subseteq\mathcal{H}$ such that both $\phi_n\to \phi$ and $A_n\phi_n\to\psi$, and we have the similar characterization of $\Gamma_\infty^*$. We can now present some basic properties of these subspaces.
\begin{lemma}\label{GraphLemma}
Let $\{A_n\}_{n=1}^\infty$ be a sequence of densely defined and closed operators and let $A$ be an operator with the same properties as the $A_n$'s.
\begin{itemize}
	\item[(a)] If $\operatorname{Gr}(A)\subseteq\Gamma_\infty$ then $\Gamma_\infty^*\subseteq\operatorname{Gr}(A^*)$.
	\item[(b)] If $\operatorname{Gr}(A)\subseteq\Gamma_\infty$ and $\operatorname{Gr}(A^*)\subseteq\Gamma_\infty^*$ then $\operatorname{Gr}(A)=\Gamma_\infty$ and $\operatorname{Gr}(A^*)=\Gamma_\infty^*$
	\item[(c)] If moreover the $A_n$'s are symmetric and $A$ is self-adjoint then $A=\operatorname{str.gr.lim }A_n$ if and only if $\operatorname{Gr}(A)\subseteq\Gamma_\infty$.
\end{itemize}
\end{lemma}
\begin{proof}
(a): Take $(\phi,\psi)\in\Gamma_\infty^*$ arbitrary and a corresponding sequence $\{\phi_n\}_{n=1}^\infty$ with $\phi_n\in D(A_n^*)$ so that $\phi_n\to\phi$ and $A_n^*\phi_n\to \psi$. Now for any $\eta\in D(A)$ there exists a sequence $\{\eta_n\}_{n=1}^\infty$ with $\eta_n\in D(A_n)$ so that $\eta_n\to\eta$ and $A_n\eta_n\to A\eta$. Using these sequences we see that
\[
\Indre{\phi}{A\eta}=\lim_{n\to\infty}\Indre{\phi_n}{A_n\eta_n}=\lim_{n\to\infty}\Indre{A_n^*\phi_n}{\eta_n}=\Indre{\psi}{\eta}
\]
proving that $\phi\in D(A^*)$ and $A^*\phi=\psi$ as needed.

(b): This is a simple application of (a) and the fact that $T^{**}=T$ for any closed operator $T$.

(c): We need only to prove that $\operatorname{Gr}(A)\subseteq\Gamma_\infty$ implies $\Gamma_\infty\subseteq\operatorname{Gr}(A)$. This is seen by the inclusions $\Gamma_\infty\subseteq\Gamma_\infty^*$ (by symmetry of the $A_n$'s) and $\Gamma_\infty^*\subseteq\operatorname{Gr}(A^*)=\operatorname{Gr}(A)$ (by (a) and self-adjointness of $A$).
\end{proof}
The connection to convergence of the projections onto the graphs of the $A_n$'s is now given in the below proposition. It tells us that the difference between strong graph convergence and strong convergence of the sequence of graph projections is measured by the absence of strong graph convergence of the sequence of adjoint operators.
\begin{pro}\label{StrongVSGraph}
Let $\{A_n\}_{n=1}^\infty$ be a sequence of densely defined and closed operators and let $A$ be an operator with the same properties as the $A_n$'s. Denote by $P_n$ and $P$ the orthogonal projections in $\mathcal{H}\oplus\mathcal{H}$ onto $\operatorname{Gr}(A_n)$ and $\operatorname{Gr}(A)$ respectively. Then $P_n\to P$ strongly if and only if both $\operatorname{Gr}(A)=\Gamma_\infty$ and $\operatorname{Gr}(A^*)=\Gamma_\infty^*$ (or equivalently if and only if both $\operatorname{Gr}(A)\subseteq\Gamma_\infty$ and $\operatorname{Gr}(A^*)\subseteq\Gamma_\infty^*$, cf. Lemma \ref{GraphLemma}(b) ).
\end{pro}
\begin{proof}
We will use the standard fact, see for example \cite{GG} Theorem 12.5, that
\begin{equation}\label{Decomposition}
\mathcal{H}\oplus\mathcal{H}=\operatorname{Gr}(T)\oplus W\operatorname{Gr}(T^*)
\end{equation}
for any densely defined and closed operator $T$ on $\mathcal{H}$ where the sum is orthogonal and $W$ is the unitary map $(\phi,\psi)\mapsto(-\psi,\phi)$.

Now if $P_n\to P$ strongly then $\operatorname{Gr}(A)=\Gamma_\infty$ by Lemma \ref{SubSpaceLemma}(b). Also, $1-P_n\to1-P$ strongly so that similarly $W\operatorname{Gr}(A_n^*)\to W\operatorname{Gr}(A^*)$ strongly by the decomposition (\ref{Decomposition}). It is an easy exercise to check that this is equivalent to $\operatorname{Gr}(A^*)=\Gamma_\infty^*$.

If, on the other hand, $\operatorname{Gr}(A)=\Gamma_\infty$ and $\operatorname{Gr}(A^*)=\Gamma_\infty^*$ then also $W\operatorname{Gr}(A_n^*)\to W\operatorname{Gr}(A^*)$ strongly. Using this we get by Lemma \ref{SubSpaceLemma}(a) that $P_nx\to Px$ for any $x\in\operatorname{Gr}(A)$ and, by using additionally (\ref{Decomposition}), $(1-P_n)y\to(1-P)y$ for any $y\in W\operatorname{Gr}(A^*)$. Combining these convergences and (\ref{Decomposition}) we conclude that $P_nz\to Pz$ for any $z\in \mathcal{H}\oplus\mathcal{H}$.
\end{proof}
To conclude this technical section we include a result on strong resolvent convergence of self-adjoint operators together with some observations. This result is well established, cf. \cite{JPS} Lemma 28 (and \cite{RS1} Theorem VIII.26 for a partial result). In the formulation below \cite{JPS} states and proves the equivalence of (i), (ii) and (iv) and \cite{RS1} that of (i) and (ii). Meanwhile, both proofs are sufficient to include also (iii) to these lists.
\begin{thm}\label{SAResult}
	Let $\{B_n\}_{n=1}^\infty$ be a sequence of self-adjoint operators on a Hilbert space $\mathcal{H}$ and $B$ another self-adjoint operator on $\mathcal{H}$. Let further $Q_n$ and $Q$ be the orthogonal projections onto $\operatorname{Gr}(B_n)$ and $\operatorname{Gr}(B)$ respectively and denote by $\Gamma_\infty^B$ the strong limit of $\{\operatorname{Gr}(B_n)\}_{n=1}^\infty$. The following statements are equivalent:
	\begin{itemize}
		\item[(i)] $B_n\to B$ in the strong resolvent sense,
		\item[(ii)] $B=\operatorname{str.gr.lim }B_n$ (i.e. $\operatorname{Gr}(B)=\Gamma_\infty^B$),
		\item[(iii)] $\operatorname{Gr}(B)\subseteq\Gamma_\infty^B$,
		\item[(iv)] $Q_n\to Q$ strongly.
	\end{itemize}
\end{thm}
\begin{proof}
	When using the self-adjointness of the operators, the equivalence between (ii) and (iii) is Lemma \ref{GraphLemma}(c) and the equivalence between (ii) and (iv) is Proposition \ref{StrongVSGraph}.
	
	Suppose $B_n\to B$ in the strong resolvent sense and let $\phi\in D(B)$ be arbitrary. By self-adjointness the relation $\psi=(B_n+i)\phi_n=(B+i)\phi$ defines besides the $\psi\in\mathcal{H}$ also for each $n$ a $\phi_n\in D(B_n)$. Moreover,
	\begin{align*}
	\norm{(\phi_n,B_n\phi_n)-(\phi,B\phi)}^2&=\norm{\phi_n-\phi}^2+\norm{B_n\phi_n-B\phi}^2=\norm{\phi_n-\phi}^2+\norm{i\phi-i\phi_n}^2
	\\
	&=2\norm{(B_n+i)^{-1}\psi-(B+i)^{-1}\psi}^2\longrightarrow 0
	\end{align*}
	which proves that (i) implies (iii).
	
	Suppose finally that $\operatorname{Gr}(B)\subseteq\Gamma_\infty^B$ and let $\psi\in\mathcal{H}$ be arbitrary. Since $\mathcal{H}=R(B+i)$ we have $\psi=(B+i)\phi$ for some $\phi\in D(B)$ and by the assumption there exists a sequence $\{\phi_n\}_{n=1}^\infty\subseteq\mathcal{H}$ so that $(\phi_n,B_n\phi_n)\to(\phi,B\phi)$. Hence,
	\[
	[(B_n+i)^{-1}-(B+i)^{-1}]\psi=(B_n+i)^{-1}[(B+i)\phi-(B_n+i)\phi_n]-\phi+\phi_n\longrightarrow0
	\]
	where we use the fact that $\norm{(B_n+i)^{-1}}\leq1$ for all $n$. This proves that (iii) implies (i) and thus the full theorem.
\end{proof}
We observe that though (ii)-(iv) in Theorem \ref{SAResult} are equivalent for sequences of self-adjoint operators, Proposition \ref{StrongVSGraph} tells us that (iii) follows from (ii) and (iv) respectively even when assuming only that the $B_n$'s and $B$ are densely defined and closed. Moreover, (iii) is a consequence of pointwise convergence on a common core of the sequence and is thus easy to verify for for example differential operators, see Proposition \ref{CommonCore} and Example \ref{DiffEx}. Thus, the first question we examine in Section \ref{sec3} below will be the following: If we in the set-up from Section \ref{sec1} impose the condition (iii) on the sequence $\{A_n\}_{n=1}^\infty$ what more do we need in order for it to hold for the sequence $\{B_n\}_{n=1}^\infty$ of extensions (thus yielding strong resolvent convergence)?
\section{Main results}\label{sec3}
From the previous section we obtain two candidates for convergence type to impose on the $A_n$'s in (\ref{result}): Strong graph convergence and strong convergence of graph projections. That these are actually both natural choices is illuminated by Theorem \ref{SAResult} which states that for sequences of self-adjoint operators each of them is equivalent to strong resolvent convergence -- exactly the convergence type we seek! Throughout this section we use the following conventions: Let $\{A_n\}_{n=1}^\infty$, $\{B_n\}_{n=1}^\infty$, $\{U_n\}_{n=1}^\infty$, $A$, $B$ and $U$ be as in the beginning
of Section \ref{sec1}. Let $\Gamma_\infty$ be the strong limit of $\{\operatorname{Gr}(A_n)\}_{n=1}^\infty$ and $\Gamma_\infty^*$ the strong limit of $\{\operatorname{Gr}(A_n^*)\}_{n=1}^\infty$. Denote by $P_n$ and $P$ the orthogonal projections in $\mathcal{H}\oplus\mathcal{H}$ onto $\operatorname{Gr}(A_n)$ and $\operatorname{Gr}(A)$ respectively.

The answer to the question closing Section \ref{sec2} is straightforward and given below in Corollary \ref{SAResultCor}, and in applications it can be useful even in this raw form.
\begin{corollary}[to Theorem \ref{SAResult}]\label{SAResultCor}
Consider operators $A_n$, $A$, $B_n$ and $B$ as described in Section \ref{sec1} and suppose $\operatorname{Gr}(A)\subseteq\Gamma_\infty$. If moreover for every pair (or, equivalently, for $k$ linearly independent pairs) $(\phi,B\phi)$ from the orthogonal complement of $\operatorname{Gr}(A)$ inside the Hilbert space $\operatorname{Gr}(B)$ there exists a sequence $\{\phi_n\}_{n=1}^\infty\subseteq\mathcal{H}$ so that $\phi_n\in D(B_n)$ for all $n$ and $(\phi_n,B_n\phi_n)\to(\phi,B\phi)$ then $B_n\to B$ in the strong resolvent sense.
\end{corollary}
\begin{proof}
Denoting the strong limit of $\{\operatorname{Gr}(B_n)\}_{n=1}^\infty$ by $\Gamma_\infty^B$ it is basically the assumption above that the orthogonal complement of $\operatorname{Gr}(A)$ inside $\operatorname{Gr}(B)$ is contained in $\Gamma_\infty^B$. Moreover, $\operatorname{Gr}(A)\subseteq\Gamma_\infty\subseteq\Gamma_\infty^B$ since all the $B_n$'s are extensions of the $A_n$'s. This concludes the proof. For the fact that it suffices to consider $k$ linearly independent pairs, see the first couple of lines of the proof of Theorem \ref{MainResult}.
\end{proof}
For the remaining part of this section we formulate and prove results like (\ref{result}) with the different notions of convergence of the $A_n$'s introduced above. For this it will be essential to have at our disposal the following characterization of strong convergence of the $U_n$'s defining the self-adjoint extensions of the $A_n$'s.
\begin{lemma}\label{ULemma}
Consider the $U_n$'s and the $U$ described in Section \ref{sec1}. We have $U_n\to U$ strongly if and only if the following statement is true:
\begin{equation}\label{UCondition}
\begin{gathered}
\text{For each $\phi_+\in\mathcal{H}_+$there exists a sequence $\{\phi_+^n\}_{n=1}^\infty\subseteq\mathcal{H}$ so that}
\\ \text{$\phi_+^n\in\mathcal{H}_+^n$ for all $n$ and $(\phi_+^n,U_n\phi_+^n)\to(\phi_+,U\phi_+)$.}
\end{gathered}
\end{equation}
\end{lemma}
Note that the condition (\ref{UCondition}) actually says that the strong limit of the graphs of the $U_n$'s considered as operators only on $\mathcal{H}_+^n$ contains the corresponding graph of $U$.
\begin{proof}[of Lemma \ref{ULemma}]
Observe firstly that if $\psi_n\to\psi$ then the inequalities
\[
\norm{U_n\psi_n-U\psi}\leq\norm{\psi_n-\psi}+\norm{U_n\psi-U\psi}\leq 2\norm{\psi_n-\psi}+\norm{U_n\psi_n-U\psi}
\]
show that 
\begin{equation}\label{UConvCond}
\text{$U_n\psi_n\to U\psi$\;\;\; if and only if\;\;\; $U_n\psi\to U\psi$.}
\end{equation}

For each $n$ denote by $P_n$ the orthogonal projection onto $\mathcal{H}_+^n$ and by $P$ the orthogonal projection onto $\mathcal{H}_+$. Assume $U_n\to U$ strongly. Then, for any $\phi_+\in\mathcal{H}_+$ and $\psi\in\mathcal{H}$, we have
\[
\Indre{P_n\phi_+}{\psi}=\Indre{U_n^*U_n\phi_+}{\psi}=\Indre{U_n\phi_+}{U_n\psi}\longrightarrow\Indre{U\phi_+}{U\psi}=\Indre{P\phi_+}{\psi}=\Indre{\phi_+}{\psi}
\]
so that $P_n\phi_+\to\phi_+$ weakly in $\mathcal{H}$. As further
\[
\norm{\phi_+}\leq\liminf_{n\to\infty}\norm{P_n\phi_+}\leq\limsup_{n\to\infty}\norm{P_n\phi_+}\leq\norm{\phi_+}
\]
by lower semi-continuity of the norm it is apparent that additionally $\norm{P_n\phi_+}\to\norm{\phi_+}$ and consequently $P_n\phi_+\to\phi_+$ with respect to the Hilbert space norm. We claim that letting $\{\phi_+^n\}_{n=1}^\infty:=\{P_n\phi_+\}_{n=1}^\infty$ for each $\phi_+\in\mathcal{H}_+$ verifies (\ref{UCondition}). Indeed, since then $\phi_+^n\to\phi_+$, the strong convergence $U_n\to U$ and (\ref{UConvCond}) yield the desired conclusion.

Suppose now that (\ref{UCondition}) is satisfied. For any $\phi_+\in\mathcal{H}_+$ we can choose the sequence from this condition and (\ref{UConvCond}) implies that $U_n\phi_+\to U\phi_+$. For proving convergence of $U_n\psi$ for $\psi\in[\mathcal{H}_+]^\perp$ fix such $\psi$ and consider an orthonormal basis $\{\phi_{+,1},\dots,\phi_{+,k}\}$ of $\mathcal{H}_+$. By (\ref{UCondition}) there exist sequences $\{\phi_{+,1}^n\}_{n=1}^\infty,\dots,\{\phi_{+,k}^n\}_{n=1}^\infty\subseteq\mathcal{H}$ with $\phi_{+,\ell}^n\in\mathcal{H}_+^n$ for all $n$ and $\ell=1,\dots,k$ and $\phi_{+,\ell}^n\to\phi_{+,\ell}$ for all $\ell$. Now by applying the Gram-Schmidt process to $\{\phi_{+,1}^n,\dots,\phi_{+,k}^n\}$ for each $n$ we obtain new sequences $\{\widetilde{\phi}_{+,1}^n\}_{n=1}^\infty,\dots,\{\widetilde{\phi}_{+,k}^n\}_{n=1}^\infty\subseteq\mathcal{H}$ with $\{\widetilde{\phi}_{+,1}^n,\dots,\widetilde{\phi}_{+,k}^n\}$ an orthonormal basis of $\mathcal{H}_+^n$ for sufficiently large $n$. Induction in $\ell$ shows that also $\widetilde{\phi}_{+,\ell}^n\to\phi_{+,\ell}$ for all $\ell$. Consequently,
\[
[\mathcal{H}_+^n]^\perp\ni\psi_n:=\psi-\sum_{\ell=1}^{k}\Indre{\widetilde{\phi}_{+,\ell}^n\;}{\psi}\widetilde{\phi}_{+,\ell}^n\longrightarrow\psi
\]
and, since $U_n\psi_n=0=U\psi$ for large $n$, a final application of (\ref{UConvCond}) proves that $U_n\psi\to U\psi$.
\end{proof}
\begin{remark}\label{Infinity}
Lemma \ref{ULemma} is actually the main reason why we assume that the deficiency indices of the $A_n$'s are finite, since then we can simply restate the condition (\ref{UCondition}) as strong convergence of the $U_n$'s -- which is exactly the kind of formulation we seek. If one, in the case of infinite deficiency indices, replaces "$U_n\to U$ strongly" with (\ref{UCondition}) then the remaining results of this note in Theorem \ref{MainResult} and Corollaries \ref{Cor1} and \ref{Cor2} indeed remain valid. One can realize that these conditions are truly different in the infinite case by taking the $U_n$'s and the $U$ to be projections and recalling the content of Remark \ref{StrictlyStronger}.
\end{remark}
While this description of strong convergence of the $U_n$'s not at first sight simplifies things, the fact that it is so closely related to the definition of strong graph convergence will help us apply our theory from Section \ref{sec2} via Theorem \ref{SAResult}. With this, we are now in a position to state and prove the main theoretical statement of this note.
\begin{thm}\label{MainResult}
Let $\{A_n\}_{n=1}^\infty$, $\{B_n\}_{n=1}^\infty$, $\{U_n\}_{n=1}^\infty$, $A$, $B$ and $U$ be as in the beginning
of Section \ref{sec1}. Let $\Gamma_\infty$ be the strong limit of $\{\operatorname{Gr}(A_n)\}_{n=1}^\infty$, and denote by $P_n$ and $P$ the orthogonal projections in $\mathcal{H}\oplus\mathcal{H}$ onto $\operatorname{Gr}(A_n)$ and $\operatorname{Gr}(A)$ respectively. Then the following holds:
\begin{itemize}
\item[(a)] If $\operatorname{Gr}(A)\subseteq\Gamma_\infty$ and $U_n\to U$ strongly then $B_n\to B$ in the strong resolvent sense and $P_n\to P$ strongly.
\item[(b)] If $B_n\to B$ in the strong resolvent sense and $P_n\to P$ strongly then $U_n\to U$ strongly.
\end{itemize}
\end{thm}
\begin{proof}
(a): Recall that, cf. \cite{RS2} X.1,
\[
\operatorname{Gr}(B)=\operatorname{Gr}(A)\oplus\Set{(\phi_++U\phi_+,i\phi_+-iU\phi_+)\given\phi_+\in\mathcal{H}_+},
\]
where the sum is orthogonal, from which the $k$-dimensional orthogonal complement of $\operatorname{Gr}(A)$ in $\operatorname{Gr}(B)$ is apparent. Applying Lemma \ref{ULemma} we can for any $\phi_+\in\mathcal{H}_+$ find $\{\phi_+^n\}_{n=1}^\infty\subseteq\mathcal{H}$ so that $\phi_+^n\in\mathcal{H}_+^n$ for all $n$ and
\[
(\phi_+^n+U_n\phi_+^n,i\phi_+^n-iU_n\phi_+^n)\longrightarrow(\phi_++U\phi_+,i\phi_+-iU\phi_+).
\]
Hence, Corollary \ref{SAResultCor} implies $B_n\to B$ in the strong resolvent sense. Likewise we have also (cf. \cite{RS2} X.1)
\[
\operatorname{Gr}(A^*)=\operatorname{Gr}(A)\oplus\Set{(\phi_+,i\phi_+)\given\phi_+\in\mathcal{H}_+}\oplus\Set{(U\phi_+,-iU\phi_+)\given\phi_+\in\mathcal{H}_+}
\]
and a similar application of Lemma \ref{ULemma} tells us that $\operatorname{Gr}(A^*)\subseteq\Gamma_\infty^*$ (the strong limit of $\{\operatorname{Gr}(A_n^*)\}_{n=1}^\infty$). By invoking Proposition \ref{StrongVSGraph} we get thus additionally $P_n\to P$ strongly.

(b): We note that by Theorem \ref{SAResult} (and using the notation herein) we have $Q_n\to Q$ strongly, and consequently $Q_n-P_n\to Q-P$ strongly. Now $Q_n-P_n$ is the orthogonal projection onto the orthogonal complement of $\operatorname{Gr}(A_n)$ inside $\operatorname{Gr}(B_n)$ and similarly for $Q-P$. But we have just seen in the proof of (a) that these are exactly
\[
\Set{(\phi_+^n+U_n\phi_+^n,i\phi_+^n-iU_n\phi_+^n)\given\phi_+^n\in\mathcal{H}_+^n}\quad\text{and}\quad\Set{(\phi_++U\phi_+,i\phi_+-iU\phi_+)\given\phi_+\in\mathcal{H}_+}
\]
respectively. Hence, Lemma \ref{SubSpaceLemma}(b) tells us that for each $\phi_+\in\mathcal{H}_+$ there exists a sequence $\{\phi_+^n\}_{n=1}^\infty\subseteq\mathcal{H}$ so that $\phi_+^n\in\mathcal{H}_+^n$ for all $n$ and
\[
(\phi_+^n+U_n\phi_+^n,i\phi_+^n-iU_n\phi_+^n)\longrightarrow(\phi_++U\phi_+,i\phi_+-iU\phi_+).
\]
By taking linear combinations of the entries we see that for this sequence $\phi_+^n\to\phi_+$ and $U_n\phi_+^n\to U\phi_+$, and to wrap things up Lemma \ref{ULemma} yields the claimed strong convergence of the $U_n$'s towards $U$ as needed.
\end{proof}
\begin{remark}
We present here a more transparent way of proving $B_n\to B$ in the strong resolvent sense in Theorem \ref{MainResult}(a) than the one presented above which in particular avoids the use of Corollary \ref{SAResultCor} and hence of Theorem \ref{SAResult}.

Define the subspace $V:=\Set{\phi_++U\phi_+\given\phi_+\in\mathcal{H}_+}$ in $\mathcal{H}$ and write
\[
\mathcal{H}=R(B+i)=R(A+i)+R(B\vert_V+i).
\]
Now since we assume $\operatorname{Gr}(A)\subseteq\Gamma_\infty$ the convergence of $(B_n+i)^{-1}$ towards $(B+i)^{-1}$ on $R(A+i)$ is proved as in (iii)$\Rightarrow$(i) in Theorem \ref{SAResult}. Notice then that
\[
(B+i)(\phi_++U\phi_+)=2i\phi_+\qquad\text{and}\qquad(B_n+i)(\phi_+^n+U_n\phi_+^n)=2i\phi_+^n
\]
for any $\phi_+\in\mathcal{H}_+$ and $\phi_+^n\in\mathcal{H}_+^n$. This proves that $R(B\vert_V+i)=\mathcal{H}_+$, and for each $\phi_+\in\mathcal{H}_+$ we can use Lemma \ref{ULemma} to find a sequence $\{\phi_+^n\}_{n=1}^\infty\subseteq\mathcal{H}$ so that $\phi_+^n\in\mathcal{H}_+^n$ for each $n$ and
\begin{align*}
\norm{(B_n+i)^{-1}\phi_+-(B+&i)^{-1}\phi_+}
\\
&\leq\norm{(B_n+i)^{-1}\phi_+-(B_n+i)^{-1}\phi_+^n}+\norm{(B_n+i)^{-1}\phi_+^n-(B+i)^{-1}\phi_+}
\\
&\leq\norm{\phi_+-\phi_+^n}+\frac{1}{2}\norm{(\phi_+^n+U_n\phi_+^n)-(\phi_++U\phi_+)}\longrightarrow 0.
\end{align*}
\end{remark}
We can now use Theorem \ref{MainResult} to prove various statements of the form (\ref{result}). Taking $A_n\to A$ to be in terms of strong convergence of the orthogonal projections onto the graphs, i.e. $P_n\to P$ strongly, we have also $\operatorname{Gr}(A)\subseteq\Gamma_\infty$ due to Proposition \ref{StrongVSGraph}, and thus we get a particularly clean statement.
\begin{corollary}\label{Cor1}
Consider the set-up in Theorem \ref{MainResult} and suppose $P_n\to P$ strongly. Then $B_n\to B$ in the strong resolvent sense if and only if $U_n\to U$ strongly.\hfill$\square$
\end{corollary}
The downside of Corollary \ref{Cor1} is, however, that the condition $P_n\to P$ strongly is often not easy to verify in concrete cases. Another approach is to assume the convergence of the $A_n$'s only in the sense that $\operatorname{Gr}(A)\subseteq\Gamma_\infty$. We note that this is a strictly weaker notion of convergence than strong convergence of the graph projections, so one cannot expect the implications of this assumption to be as strong as the equivalence between strong convergence of the $B_n$'s and of the $U_n$'s in Corollary \ref{Cor1}. Another application of Proposition \ref{StrongVSGraph} yields:
\begin{corollary}\label{Cor2}
Consider the set-up in Theorem \ref{MainResult} and suppose $\operatorname{Gr}(A)\subseteq\Gamma_\infty$. Then $U_n\to U$ strongly if and only if both $B_n\to B$ in the strong resolvent sense and $\operatorname{Gr}(A^*)\subseteq\Gamma_\infty^*$.\hfill$\square$
\end{corollary}
An obvious question now arises: Is this the best we can do? In particular we can in the light of Corollary \ref{Cor1} ask whether the condition $\operatorname{Gr}(A^*)\subseteq\Gamma_\infty^*$ in Corollary \ref{Cor2} is actually needed. As a matter of fact it is by the following observations.
\begin{remark}
We do not in general have the result "Suppose $\operatorname{Gr}(A)\subseteq\Gamma_\infty$. Then $U_n\to U$ strongly if and only if $B_n\to B$ in the strong resolvent sense." as the example below shows. Even changing $\operatorname{Gr}(A)\subseteq\Gamma_\infty$ to $A=\operatorname{str.gr.lim }A_n$ does not make the statement true. The backbone of the example is the extension theory for a well-studied class of operators on $L^2(\mathbb{R}^3)=\mathcal{H}$. This is treated in for example \cite{SMiQM} I.1.1 to which we refer for the details.

Let $\{y_n\}_{n=1}^\infty\subseteq\mathbb{R}^3$ be a sequence yet to be specified and define for each $n$ the operator $A_n$ to be the closure of $-\Delta$ on $C_c^\infty(\mathbb{R}^3\backslash\{y_n\})$. One can now find the deficiency subspaces
\[
\mathcal{H}_{\pm}^n=\mathbb{C}\phi_\pm^n,\qquad\phi_\pm^n(x)=\frac{e^{i\sqrt{\pm i}\abs{x-y_n}}}{4\uppi\,\abs{x-y_n}}
\]
where $\Im\sqrt{\pm i}>0$. Moreover, if one defines a self-adjoint extension $B_n$ of $A_n$ by the unitary map $U_n\colon\mathcal{H}_+^n\ni\phi_+^n\mapsto-\phi_-^n\in\mathcal{H}_-^n$ as in Section \ref{sec1} then $B_n=B$ is actually the free Laplacian $-\Delta$ defined on the Sobolev space $H^2(\mathbb{R}^3)$ \emph{independently} of $n$. Now we have the orthogonal decomposition
\[
\operatorname{Gr}(B)=\operatorname{Gr}(A_n)\oplus\mathbb{C}(\phi_+^n-\phi_-^n,i\phi_+^n+i\phi_-^n)=:\operatorname{Gr}(A_n)\oplus\mathbb{C}v_n,
\]
and consequently $\operatorname{Gr}(A_n)$ is the orthogonal complement of $\mathbb{C}v_n$ in $\operatorname{Gr}(B)$ for each $n$. Notice now that the $v_n$'s depend only on the $y_n$'s. Choosing $y_n$ so that $\abs{y_n}\to\infty$ it is not difficult to realize that the sequences $\{\phi_\pm^n\}_{n=1}^\infty$ converge weakly towards $0$ in $L^2(\mathbb{R}^3)$: This follows from the fact that they are translations of a fixed $L^2$-function. With such sequence of $y_n$'s we get thus
\[
\Indre{(\phi,\psi)}{v_n}=\Indre{\phi}{\phi_+^n}-\Indre{\phi}{\phi_-^n}+i\Indre{\psi}{\phi_+^n}+i\Indre{\psi}{\phi_-^n}\longrightarrow0
\]
for all $(\phi,\psi)\in\mathcal{H}\oplus\mathcal{H}$, i.e. $v_n\to 0$ weakly in $\mathcal{H}\oplus\mathcal{H}$ and hence in $\operatorname{Gr}(B)$.

We observe from the above facts that by choosing a sequence of $y_n$'s which is a fixed $y_n=y_0$ for $n$ odd and with $\{y_{2n}\}_{n=1}^\infty$ unbounded we can make the sequence $\{\operatorname{Gr}(A_n)\}_{n=1}^\infty$ of subspaces of the Hilbert space $\operatorname{Gr}(B)$ into a sequence like $\{V_n\}_{n=1}^\infty$ in Remark \ref{StrictlyStronger}. Consequently, the operator $A=A_1$ is the strong graph limit of the $A_n$'s (and of course $B_n\to B$), but the orthogonal projections onto the graphs $\operatorname{Gr}(A_n)$ do not converge strongly towards the orthogonal projection onto $\operatorname{Gr}(A)$, and hence Theorem \ref{MainResult}(a) tells us that we cannot have $U_n\to U$ strongly. Alternatively this can be checked more directly by using Lemma \ref{ULemma}.
\end{remark}
We conclude by proving a simple requirement for having $\operatorname{Gr}(A)\subseteq\Gamma_\infty$, thus providing a procedure for checking the assumptions in Corollary \ref{Cor2}. Recall that a core for a closed operator $A$ is a subspace of $D(A)$ satisfying that the restriction of $A$ to this has closure $A$. We obtain now:
\begin{pro}\label{CommonCore}
Assume that $\mathcal{D}$ is a common core for $A$ and all $A_n$'s. If $A_n\phi\to A\phi$ for all $\phi\in\mathcal{D}$ then $\operatorname{Gr}(A)\subseteq\Gamma_\infty$.
\end{pro}
\begin{proof}
The assumption tells us that $\Gamma_\mathcal{D}:=\Set{(\phi,A\phi)\given\phi\in\mathcal{D}}\subseteq\Gamma_\infty$. Thus, if we argue that $\Gamma_\infty$ is closed, we have also $\operatorname{Gr}(A)=\overline{\Gamma_\mathcal{D}}\subseteq\Gamma_\infty$. But closedness is a general property of any strong limit of subspaces by the following argument:

Let $\{V_n\}_{n=1}^\infty$ be any sequence of subspaces of a Hilbert space $\mathcal{H}$ and denote as usual its strong limit by $V_\infty$. If we consider an arbitrary convergent sequence $\{x_k\}_{k=1}^\infty\subseteq V_\infty$ with limit $x_0$ then we need only to find a sequence $\{\widetilde{x}_n\}_{n=1}^\infty\subseteq\mathcal{H}$ with $\widetilde{x}_n\in V_n$ for all $n$ such that $\widetilde{x}_n\to x_0$ in order to obtain $x_0\in V_\infty$ and hence prove that $V_\infty$ is closed. We now construct such sequence. Firstly we choose for each $k$ a sequence $\{x_n^k\}_{n=1}^\infty$ with $x_n^k\in V_n$ for all $n$ and $x_n^k\to x_k$, and then we take natural numbers $N_1<N_2<N_3<\cdots$ so that $\norm{x_n^k-x_k}<1/k$ for all $n\geq N_k$. Defining $\widetilde{x}_n:=x_n^1$ for $n=1,2,\dots,N_2-1$; $\widetilde{x}_n:=x_n^2$ for $n=N_2,\dots,N_3-1$ and generally $\widetilde{x}_n:=x_n^k$ for $n=N_k,\dots,N_{k+1}-1$ one can check using the triangular inequality that this is indeed a sequence with the properties we seek.
\end{proof}
\begin{example}\label{DiffEx}
To make things even more concrete than requiring pointwise convergence of the $A_n$'s on a common core, we can ask what this means for differential operators like those in Example \ref{Motivation}. To simplify things let us consider a sequence of Schrödinger operators -- that is, the $A_n$'s are the closures of $-\Delta+\Phi_n$ defined on $C_c^\infty(\Omega)\subseteq L^2(\Omega)$ for some open set $\Omega\subseteq\mathbb{R}^d$ and some potentials $\Phi_n$ (say, real-valued and continuous) on this set. Hence, $C_c^\infty(\Omega)$ is a common core for the $A_n$'s and also for $A=-\Delta+\Phi$ if we define this in the same manner. Now, for any $\phi\in C_c^\infty(\Omega)$,
\[
\norm{A_n\phi-A\phi}^2=\norm{\Phi_n\phi-\Phi\phi}^2=\int_\Omega\abs{\phi}^2\abs{\Phi_n-\Phi}^2\,dx\leq\norm{\phi}_\infty^2\int_{\operatorname{supp}\phi}\abs{\Phi_n-\Phi}^2\,dx
\]
where $\norm{\,\cdot\,}_\infty$ is the supremum norm. Now if $\Phi_n\to\Phi$ in $L^2_{\operatorname{loc}}(\Omega)$ then we conclude that $A_n\phi\to A\phi$ for all $\phi\in C_c^\infty(\Omega)$. If, on the other hand, we assume the latter, then we see that $\Phi_n\to\Phi$ in $L^2(K)$ for any compact subset $K\subseteq\Omega$ by choosing $\phi\equiv1$ on $K$, i.e. we get $\Phi_n\to\Phi$ in $L^2_{\operatorname{loc}}(\Omega)$. Being able to consider only local $L^2$-convergence is often desirable if one deals for example with potentials with singularities.
\end{example}
\section*{Acknowledgements}
This work was supported in parts by the VILLUM Foundation grant no. 10059. I would like to thank my PhD advisor Jan Philip Solovej for initiating this cute little project as well as for his always committed and insightful guidance. Also, I thank Johannes Agerskov for proofreading and for his useful suggestions.
\section*{Comments}
This version of the article has been accepted for publication, after peer review but is not the Version of Record and does not reflect post-acceptance improvements, or any corrections. The Version of Record is available online at \href{https://doi.org/10.1007/s00023-023-01397-9}{https://doi.org/10.1007/s00023-023-01397-9}.
\bibliographystyle{chronoplainnm}
\bibliography{ref.bib}
\end{document}